\documentstyle[11pt]{article}

\input epsf

\begin{document}
\thispagestyle{empty}

\begin{center}
\null\vspace{-1cm}

{\bf EXTENDED SUPER-KP HIERARCHIES AND GENERALIZED STRING EQUATIONS}\\[0pt]
\vspace{1.5cm} B. Maaroufi, M. Nazah, M.B. Sedra\footnote{%
Regular Associate of the Abdus Salam ICTP. Corresponding author:
mysedra@yahoo.fr}\\[0pt]
{\it Laboratoire de Physique Th\'eorique et Appliqu\'ee LPTA,\\[0pt]
Universite Ibn Tofail, Facult\'e des Sciences, D\'epartement de Physique,\\[%
0pt]
B.P. 133, Kenitra, Morocco\\[0pt]
and\\[0pt]
The Abdus Salam International Centre for Theoretical Physics, Trieste, Italy.%
}\\[0pt]
\end{center}

\vspace{0.5cm}

\centerline{\bf Abstract} \baselineskip=18pt \bigskip

We propose a consistently algebraic formulation of the extended KP
(supersymmetric) integrable-hierarchy systems. We exploit the results
already established in [14] and which consist in a framework suspected to
unify in a fascinating way all the possible supersymmetric KP-hierarchies
and then their underlying supergravity theories . This construction leads
among other to built explicit non standard integrable Lax evolution
equations suspected to reduce to the well known KP integrable equation. We
present also a contribution of our construction to the subject of string
equation and solitons. Other algebraic properties are also presented.
\newpage

\section{Introduction}

An interesting subject which have been studied recently from different point
of views deals with the field of non linear integrable systems and their
higher and lower spin extensions [1, 2]. These are exactly solvable models
exhibiting a very rich structure in lower dimensions and are involved in
many areas of mathematical physics. One recall for instance the two
dimensional Toda(Liouville) fields theories [3, 2] and the KdV and KP
hierarchies models [4, 5], both in the bosonic as well as in the
supersymmetric case.

Non linear integrable models, are associated to systems of non linear
differential equations which can be solved exactly. Mathematically these
models have become more fascinating by introducing some new concepts such as
the infinite dimensional Lie (super) algebras.[6], Kac- Moody algebras [7],
W-algebras [1, 2], quantum groups [8] and the theory of formal
pseudodifferential operators [4, 5]. From the physical point of view only
recently after the discovery of the connection between the generalized KdV
hierarchies and the matrix models formulation of two dimensional gravity
[9], has there been a great progress in the study of these systems. Wile the
bosonic non linear integrable models have been studied quite extensively,
not much is known, in general, about the supersymmetric extension.

The most widely studied supersymmetric integrable models which have many
interesting properties is the super-KdV (sKdV) system [10]. Note by the way
that a super-integrable model which exhibits a very rich structure and which
in fact leads to the sKdV system upon reduction is the supersymmetric KP
equation of Manin and Radul [11]. Another interesting integrable system
which have been also generalized recently to the supersymmetric case is the
two boson system [12]. This model known also as the dispersive long water
wave equation has a very rich structure since it is :1- Tri Hamiltonian , 2-
has a non standard Lax representation and 3- Reduces to well known
integrable systems under appropriate reduction.

Non linear integrable models are also known to connect with the recent
subject of string theory and its various applications [9]. This deep
connection is established through solitons which characterise in some sense
non linear physical models [13]. For a review, recall that in the late of
60-ies and early 70-ies a large group of physicists developed the very
beautiful theory of the bosonic quantum strings. They used in this issue
standard operator quantisation, decomposing fields in the Fourier series and
replacing c-numbers by operators with standard canonical commutators. This
program was effectively realized for the ?zero-loop? described by the
Riemann surfaces of the zero genus. The underlying Virasoro algebra
(conformal symmetry) and its representations are known to play an important
role. This program stopped because nobody was able to quantize fields in
such a way for non zero genus. In early 80-ies Polyakov solved the problem
of quantization of bosonic strings using the functional path integral. Later
on many physicist focussed to complete much more this program by using the
analytical construction of the Soliton theory in Riemann surfaces. Recently
a modern terminology was introduced to describe the connection between the
theory of solitons based on pseudo-differential operators and string theory.
This consist in what we know as ?string equation? which means exactly the
following pseudo-differential equation:
\begin{equation}
[L , A ]=1
\end{equation}
String equation appeared in 1989-90 years after the pioneering works of
Gross, Migdal, Brezin, Kazakov, Douglas, Shanker, David and many others in
matrix models formulation of gravity theories [9]. In this work ; we propose
a consistently algebraic formulation of the extended KP integrable hierarchy
systems both in the bosonic and supersymmetric cases. We exploit the results
already established in [14] and which consist in a framework suspected to
unify in a fascinating way all the possible supersymmetric KP-hierarchies
and then their underlying supergravity theories . This construction leads
among other to built explicit non standard integrable Lax evolution
equations suspected to reduce to the well known KP integrable equation. We
present also a contribution of our construction to the subject of string
equation. Other algebraic properties are also presented.

\section{The Lie algebra of pseudo differential operators}

In this section, we give the general setting of the basic properties of the
tensor algebra of the bosonic W currents of conformal spin $s>1$. This is a
semi-infinite dimensional space of the infinite tensor algebra of arbitrary
integer spin fields. Recall that W currents, one of the main actors in two
dimensional conformal theory and their higher spin extensions, are analytic
fields obeying a nonlinear closed algebra. The W fields also appear in the
study of higher differential operators involved in the analysis of nonlinear
integrable models to be considered in this article. We shall start however
by defining our notations.

The two dimensional Euclidean space ${R}^{2}\cong {C}$ is parametrized by $%
z=t+ix$ and $\bar{z}=t-ix$. As a matter of convention, we set $z=z^{+}$ and $%
z=z^{-}$ so that the derivatives $\partial /\partial z$ and $\partial
/\partial \bar{z}$ are, respectively, represented by $\partial _{+}=\partial
$ and $\partial _{-}$=$\bar{\partial}$. The so(2) Lorentz representation
fields are described by one component tensors of the form $\psi _{k}(z,\bar{z%
})$ with $2k\in {Z}$. ${Z}$ is the set of relative integers. In two
dimensional conformal field theories (CFT), an interesting class of fields
is given by the set of analytic fields $\phi _{k}(z)$. These are $SO(2)\cong
U(1)$ tensor fields that obey the analyticity condition $\partial _{-}\phi
_{k}(z)=0$. In this case the conformal spin $k$ coincides with the conformal
dimension. Note that under a $U(1)$ global transformation of parameter $%
\theta $, the object $z^{\pm }$, $\partial _{\pm }$ and $\phi _{k}(z)$
transform as
\begin{equation}
z^{\pm \prime }=e^{\mp i\theta }z^{\pm },\quad \partial _{\pm }^{\prime
}=e^{\pm i\theta }\partial _{\pm },\quad \phi _{k}^{\prime }(z)=e^{ik\theta
}\phi _{k}(z)  \label{2.1}
\end{equation}
so that $dz\partial _{z}$ and $(dz)^{k}\phi _{k}(z)$ remain invariant. In a
pure bosonic theory, which is the purpose of the present study, only integer
values of conformal spin $k$ are involved. We denote by $\Xi ^{(0,0)}$ the
tensor algebra of analytic fields of arbitrary conformal spin. This is a
completely reducible infinite dimensional SO(2) Lorentz representation
(module) that can be written as
\begin{equation}
\Xi ^{(0,0)}={\oplus }_{k\in {Z}}\Xi _{k}^{(0,0)}  \label{2.2}
\end{equation}
where the $\Xi _{k}^{(0,0)}$'s are one dimensional $SO(2)$ spin $k$
irreducible modules. The upper indices $(0,0)$ carried by the spaces
figuring in Eq.(3) are special values of general indices $(p,q)$ to be
introduced later on.

Next we introduce the space of pseudo-differential operators whose elements $%
L_{s}^{(p,q)}$ are the generalization of the well known differential Lax
operators involved in the analysis of the so called KdV-hierarchy and in
Liouville (Toda) field theories. The simplest example is given by the Hill
operator
\begin{equation}
L_{2}=\partial ^{2}+u_{2}(z)
\end{equation}
which play an important role in the study of Liouville field theory and in
the KdV equation. A natural generalization of the above relation is given by
[15]
\begin{equation}
L_{s}=\sum_{i=p}^{q}u_{s-i}(z)\partial ^{i}
\end{equation}

where $u_{s-i}(z)$ s are analytic fields of spin $(s-i)$ and where $p$ and $%
q $ with $p\leq q$ are integers that we suppose positive for the moment. We
shall refer hereafter to $p$ as the lowest degree of \ $L_{s}^{(p,q)}$ and $%
q $ as the highest degree. We consider these two features of \ Eq (5) by
setting
\begin{equation}
\deg (L_{s}^{(p,q)})=(p,q)
\end{equation}
and
\begin{equation}
\Delta (L_{s}^{(p,q)})=s
\end{equation}
as been the conformal weight. Note that the KdV operator Eq(2.3) is
discovered from Eq(2.4) as a special case by setting $s=2,$ $p=0$ and $q=2$
together with the special choice $u_{0}=1$ and $u_{1}=0$ corresponding to
special choice dealing with Lie algebraic $sl(2)$. Moreover, Eq(5) which is
well defined for $0\leq $ $p\leq q$ , may be extended to negative integers
by introducing pseudo-differential operators of type $\partial ^{-i},i\geq 1$
whose action on the fields $u_{s}(z)$ is defined as
\begin{equation}
\partial ^{-k}u_{s}(z)=\sum_{l=0}^{\infty
}(-1)^{l}c_{k+l-1}^{l}u_{s}^{(l)}\partial ^{-k-l}
\end{equation}
We have
\begin{equation}
\partial ^{k}\partial ^{-k}u_{s}(z)=u_{s}(z)  \label{3.6}
\end{equation}
As it was previously noted, a natural representation basis of nonlinear
pseudodifferential operators of spin $m$ and negative degrees $(p,q)$ is
given by
\begin{equation}
\delta _{m}^{(p,q)}(u)=\sum_{i=p}^{q}u_{m-i}(z)\partial ^{i}  \label{3.7}
\end{equation}
This configuration is useful in the study of the algebraic structure of the
spaces $\Xi _{m}^{(p,q)}$ and $\Xi ^{(p,q)}$. Note also that we can use
another representation of pseudodifferential operators, namely, the Volterra
representation. This is convenient in the derivation of the second
Hamiltonian structure of higher conformal spin integrable theories.
Moreover, ne sees that operators with negative lowest degrees $p$ and
positive highest degrees $q$ denoted by $D_{m}^{(p,q)}[u]$ split as
\begin{equation}
D_{m}^{(p,q)}\left[ u\right] =\delta _{m}^{(p,q)}(u)+d_{m}^{(p,q)}(u)
\label{3.8}
\end{equation}
More generally we have
\begin{equation}
D_{m}^{(p,q)}\left[ u\right] =D_{m}^{(p,k)}(u)+D_{m}^{(k+1,q)}(u)
\label{3.9}
\end{equation}
for any integers $p\leq k\leq q$. As a consequence, one finds that
\begin{equation}
(p,q)=(p,k)+(k+1,q)  \label{3.10}
\end{equation}
for any three integers such that $p\leq k\,<q$. Now let $\Xi _{m}^{(p,q)}$; $%
m,p,$ and $q$ integers with $q\geq p$, be the set of spin $m$ differential
operators of degrees $(p,q)$, the $\Xi _{m}^{(i,i)}$'s are one dimensional
spaces given by
\begin{equation}
\Xi _{m}^{(i,i)}=\Xi _{m-i}^{(0,0)}\otimes \partial ^{i}  \label{3.12}
\end{equation}

\section{\protect\bigskip Unified framework of bosonic KP-hierarchy}

Recall that KP hierarchies can be thought of as a dynamical system defined
on a space whose functions $u_{j}(z)$ are elements of the ring of analytic
fields. It is also defined as the universal family of isospectral
deformations of the pseudo-differential operator;
\begin{equation}
L=\partial +\sum_{i=1}^{\infty }u_{i}(z)\partial ^{1-i}
\end{equation}
satisfying the evolution equation
\begin{equation}
\frac{\partial L}{\partial t_{n}}=\left[ (L^{n})_{+},L\right],%
n=1,2,...
\end{equation}
where the subscripts + means taking purely differential part of $(L^{n}).${\
}Now; we are very interested in generalizing this standard integrable
KP-hierarchy. A natural way to do it is to focus the conformal spin
decomposition. This allows to suppose the existence of three classes of
integrable KP-hierarchies generalizing the standard ones and which are
labelled by the conformal spin quantum number $s=\pm ,0$. A part of the Lax
operator Eq() generating $\Sigma _{+}$ one can introduce two other classes
of integrable hierarchies which are described by the following Lax operators

\begin{equation}
L_{0}=u_{-1}\partial +\sum_{i=0}^{\infty }u_{i}(z)\partial ^{-i}
\end{equation}

generating the subspace $\sum_{0}$ of Lorentz scalar pseudo-operators and

\begin{equation}
L_{-}=u_{-2}\partial +\sum_{i=-1}^{\infty }u_{i}(z)\partial ^{-i-1}
\end{equation}

which belongs to $\Sigma _{-}.$ Actually, we now well known that the origin
of these three classes of integrable KP-hierarchy is traced to the fact that
there exist precisely three decompositions of $\Sigma $ into a linear sum of
two subspaces namely [14-15]
\begin{equation}
\Sigma =\oplus _{s=\pm ,0}\left( \Sigma _{s}^{+}\oplus \Sigma
_{s}^{-}\right)
\end{equation}
as given in Eq(). This provide then a unified framework in which all the
possible KP-hierarchies described by various Lax operators can be written.

\bigskip

\section{Unified framework of supersymmetric KP-hierarchy}

\subsection{The space of higher spin supersymmetric Lax operators}

The aim of this section is to describe the supersymmetric version of the
space of bosonic Lax operators introduced previously. This supersymmetric
generalization which is straightforward and natural in the fist steps,
exhibits however, some properties which are ignored in the bosonic case and
make the fermionic study more fruitful. Using the space of supersymmetric
Lax operators, one can derive the Hamiltonian structure of non linear two
dimensional super integrable models extending the bosonic Hamiltonians.

Let us first consider the ring of all analytic super fields $u_{\frac{k}{2}}(\hat{z}),~k\in {Z}$, which depend on $(1\mid 1)$ superspace coordinates
$\hat{z}=(z,\theta )$. In this super commutative ${Z}_{2}- $graded ring
R, one can define an odd super derivation $D=\partial _{\theta }+\theta
\partial $, the N=1 supercovariant derivatives which obeys the N=1
supersymmetric algebra $D^{2}=\partial $ with $\theta ^{2}=0$ and $\partial
_{\theta }=\int d\theta $.

Note that the supersymmetric G.D bracket, which we shall discuss in the
sequel, defines a Poisson bracket on the space of functional of the
superfields $u_{\frac{k}{2}}(\hat{z})$ defined on the ring $R\left[ u(\hat{z}%
)\right] $.

We define the ring $\Sigma \left[ D\right] $ of differential supersymmetric
operators as polynomials in $D$ with coefficients in $R$. using our previous
notation, one set
\begin{equation}
\Sigma [D]={\oplus }_{m\in Z}{\oplus }_{p\leq q}\Sigma _{\frac{m}{2}}^{(p,q)}[D], p,q\in {Z} 
\end{equation}
where $\Sigma _{\frac{m}{2}}^{(p,q)}[D]$ is the space of supersymmetric
operators type
\begin{equation}
\pounds _{\frac{m}{2}}^{(p,q)}[u]=\sum_{i=p}^{q}u_{\frac{m-i}{2}}(\hat{z}%
)D^{i}\quad p,q\in {Z}
\end{equation}
$\Sigma _{\frac{m}{2}}^{(p,q)}$ behaves as a $(1+q-p)$ dimensional
superspace. Note also that the ring $R$ of all graded superfields can be
decomposed as
\begin{equation}
R\equiv R^{(0,0)}:={\oplus }_{k\in Z}R_{\frac{k}{2}}^{(0,0)}
\end{equation}
where $R_{\frac{k}{2}}^{(0,0)}$ is the set of superfield $u_{\frac{k}{2}}(%
\hat{z})$ indexed by half integer conformal spin $\frac{k}{2}\in { Z+}%
\frac{1}{2}$. Thus, the one dimensional objects $u_{\frac{m-i}{2}}(\hat{z}%
)D^{i}$ are typical elements of the superspace
\begin{equation}
\Sigma _{\frac{m}{2}}^{(i,i)}=R_{\frac{m-i}{2}}^{(0,0)}.D^{i}\equiv R_{\frac{%
m-i}{2}}^{(0,0)}\otimes \Sigma _{\frac{i}{2}}^{(i,i)}
\end{equation}
which is fundamental in the construction of supersymmetric operators type
(3.2).

The expression Eq(3.4) means also that
\begin{equation}
\Sigma _{\frac{m}{2}}^{(p,q)}[D]\equiv {\oplus }_{i=p}^{q}\Sigma _{\frac{m-i}{2}}^{(i,i)} 
\end{equation}
Indeed these are important in the sense that one easly identify all objects
of the huge superspace $\Sigma $.

An element $\pounds $ of $\Sigma \lbrack D]$ is called a supersymmetric Lax
operator if it is homogeneous under the ${Z}_{2}$-grading
\begin{equation}
\left| x\right| :=\left\{
\begin{array}{c}
0\quad ,\,\,x\,\,\,\,even \\
1\quad ,\,\,x\,\,\,\,odd
\end{array}
\right. 
\end{equation}

And have the following form at order $n,n\in {N}$%
\begin{equation}
\pounds _{\frac{n}{2}}^{(0,n)}:=\sum_{i=0}^{n}u_{\frac{i}{2}}(\hat{z})D^{n-i}
\end{equation}

The homogeneity condition simply states that the ${Z}_{2}$-grading of
the N=1 superfield $u(\hat{z})$ is defined by
\begin{equation}
\left| U_{\frac{i}{2}}(\hat{z})\right| =i\,({mod}\,2)
\end{equation}
The space of supersymmetric Lax operators is refereed hereafter to as $\Sigma
_{\frac{n}{2}}^{(0,n)}$ and exhibits a dimension n+1.

\begin{enumerate}
\item  We recall that the upstairs integers (0,n) are the lowest and the
highest degrees of $\pounds $ \ and the down stair index $\frac{n}{2}$ is
the spin of $\pounds $. To define a Lie algebraic structure on the superspace
$\Sigma $ one need to introduce a graded commutator defined for two
arbitrary operators X and Y by
\begin{equation}
\left[ X,Y\right] _{i}=XY-(-)^{i}YX 
\end{equation}
Where the index i=0 or 1 refer to the commutator [,] or anticommutator \{,\}
respectively. The multiplication of operators in $\Sigma $ is defined be the
super Leibniz rule given by the following mapping
\[
D^{(l)}:R_{\frac{j}{2}}^{(0,0)}\longrightarrow \Sigma _{\frac{j+l}{2}%
}^{(p,l)}
\]
\begin{equation}
D^{(l)}\phi \left( \hat{z}\right) =\sum_{i=0}^{\infty }\left[
\begin{array}{c}
l \\
l-i
\end{array}
\right] (-)^{j(l-i)}\phi ^{(i)}D^{(l-i)}
\end{equation}
Where l is an arbitrary integer and the super binomial coefficients $\left[
\begin{array}{c}
l \\
k
\end{array}
\right] $ are defined by [..]
\begin{equation}
\left[
\begin{array}{c}
l \\
k
\end{array}
\right] =\left\{
\begin{array}{c}
0, for k\succ l and for (k,l)=(0,1){mod}2 \\
\left(
\begin{array}{c}
l \\
k
\end{array}
\right) , {otherwise }
\end{array}
\right.  
\end{equation}
The lowest degree p of the superspace $\Sigma _{\frac{j+l}{2}}^{(p,l)}$
Eq(3.12) is given by
\begin{equation}
p=\left\{
\begin{array}{c}
0, if l\geq 0 \\
-\infty  if l\leq -1
\end{array}
\right.  
\end{equation}
The symbol [x] stands for the integer part of x$\in { Z}/2$ and $\left(
\begin{array}{c}
i \\
j
\end{array}
\right) $ is the usual binomial coefficient.
\end{enumerate}

\subsubsection{The 5=2x2+1 splitting:}

Recall that there are two usual supersymmetric extensions of the standard
KP-hierarchy Eq (3.1). The first one is given by the Manin-Radul
supersymmetric KP hierarchy associated with the odd super Lax operator[11]:

\begin{enumerate}
\item
\begin{equation}
L_{MR}=D+\sum_{i\geq 1}^{\infty }u_{\frac{i+1}{2}}(\hat{z})D^{-i}
\end{equation}
\end{enumerate}

The second one is given by the supersymmetric KP-hierarchy associated with
the Figueroa-Mass- Ramos even super Lax operator [16]\bigskip
\begin{equation}
L_{FMR}=D^{2}+\sum_{i\geq 0}^{\infty }u_{\frac{i+1}{2}}(\hat{z})D^{-i}
\end{equation}
An important consequence of the choice of $L_{FMR}$, Eq(4.16), is that under
suitable choice of reduction, it reduce to the Inami-Kanno super Lax
operator describing the generalised N=2 super KdV-hierarchy [17]. Recall
also the important remark of [14], in which a unifying framework of the
previous well known super KP-hierarchies is proposed. Indeed, using the $%
6=3\times 2$ decomposition of the Lie superalgebra $\Xi $, Eq(3.6), we have
shown that there exist precisely $5=2\times 2+1$ classes of supersymmetric
KP-hierarchies. The origin of these hierarchies is traced to the fundamental
fact that there exist precisely five grading algebras:

\bigskip
\begin{eqnarray*}
g_{1} &=&\Xi _{+,\overline{0}}^{+}\oplus \Xi _{+,\overline{0}}^{-} \\
g_{2} &=&\Xi _{+,\overline{1}}^{+}\oplus \Xi _{+,\overline{1}}^{-} \\
g_{0} &=&\Xi _{0,\overline{0}}^{+}\oplus \Xi _{0,\overline{0}}^{-} \\
g_{1}^{\ast } &=&\Xi _{-,\overline{1}}^{+}\oplus \Xi _{-,\overline{1}}^{-} \\
g_{2}^{\ast } &=&\Xi _{-,\overline{0}}^{+}\oplus \Xi _{-,\overline{0}}^{-}
\end{eqnarray*}

\bigskip

\section{References}

[1] A.B.Zamolodchikov, TMP 65 (1985)1205

V.A.Fateev and A.B.Zamolodchikov, Nucl. Phys. B 280[FS18](1988)6411

[2] P. Bouwknegt and K.Schoutens, Phys. Rep. 233(1992)183 and references
therein

[3] P. Mansfield, Nucl. Phys.B 208 (1982)277, B222(1983)419

D.Olive and N.Turok, Nucl. Phys. B 257[FS14](1986)277

L. Alvarez-Gaum\'{e} and C. Gomez, Topics in Liouville Theory, CERN
preprint-Th6175/91

[4] P.D.Lax, Commun. Pure Appl. Math. 21 (1968)476 and 28 (1975)141

M. Jimbo and T. Miwa, Integrable systems in statistical mechanics, Ed. G
DAriano et al,

World Scientific 1990

L.D.Fadeev and L.A. Takhtajan, Hamiltonian methods in the Theory of solitons,

Springer Berlin 1987

[5] A.Das, Integrable models, World Scientific 1989

[6] J.E. Humphreys, Introduction to Lie algebras and representation theory,

Springer He\"{i}delberg (1972)., V. Kac, Advances in Math. 26(1977)8

[7] W.Z.Xian, Introduction to Kac-Moody algebras, world Scientific 1991

[8] V.G. Drinfeld, Quantum groups, Proc. Of the International congress of
mathematicians,

Berkeley, 1986, American Mathematical Society 798-1987

[9] For review see P.Di-Francesco, P. Ginsparg and J.Zinn-Justin, Phys. Rep.
254(1995)1

[10] P. Mathieu, J.Math. Phys. 29(1988)2499, B.A.Kupershmidt, Phys. Lett.
A102(1984)213

[11] Yu.I.Manin and A.O.Radul, Cummun. Math. Phys. 98(1985)65

[12] J.C.Brunelli and A.Das, Phys. Lett. B337, 303(1994).

J.C.Brunelli and A.Das, Int. J.Mod.Phys.A10, 4563(1995).

[13] L.A.Dickey, Soliton equations and hamiltonian systems, World Scientific
1991

A.C.Newell, Solitons in mathematics and Physics (Society for industrial and
applied

mathematics, Philadelphia 1985)

S.P.Novikov, The string equation and Solitons, hep-th/9512094

[14] M.B. Sedra, J. Math. Phys.37, 3483(1996)

[15] E.H.Saidi and M.B. Sedra, J. Math. Phys.35, 3190(1994)

[16] J.M.Figueroa-OFarrill, J.Mas and E. Ramos, Rev. Math. Phys. 3,
479(1991)

[17] T.Inami and H. Kanno, preprint YITP/K-928/91; Int.J.Mod.Phys.A7,
Supp.1A-419(1992)

[18] M.R.Douglas, Phys. Lett.B238 (1990)176
\end{document}